\documentclass[prodmode,acmtecs]{acmsmall} 

\usepackage{graphicx}
\usepackage{caption}
\usepackage{subfig}
\usepackage{tabulary}
\usepackage{lineno}

\usepackage{alltt}

\usepackage{listings}
\usepackage[cmex10]{amsmath}
\usepackage{url}

\usepackage[ruled]{algorithm2e}
\renewcommand{\algorithmcfname}{ALGORITHM}
\SetAlFnt{\small}
\SetAlCapFnt{\small}
\SetAlCapNameFnt{\small}
\SetAlCapHSkip{0pt}
\IncMargin{-\parindent}

\begin{document}

\markboth{F. Luporini et al.}{COFFEE: an Optimizing Compiler for Finite Element Local Assembly}

\title{COFFEE: an Optimizing Compiler for Finite Element Local Assembly}
\author{Fabio Luporini
\affil{Imperial College London}
Ana Lucia Varbanescu
\affil{University of Amsterdam}
Florian Rathgeber
\affil{Imperial College London}
Gheorghe-Teodor Bercea
\affil{Imperial College London}
J. Ramanujam
\affil{Louisiana State University}
David A. Ham
\affil{Imperial College London}
Paul H.J. Kelly
\affil{Imperial College London}}

\begin{abstract}
The numerical solution of partial differential equations using the
finite element method is one of the key applications of high performance
computing. Local assembly is its characteristic operation. This entails
the execution of a problem-specific kernel to numerically evaluate an
integral for each element in the discretized problem domain. Since the
domain size can be huge, executing efficient kernels is fundamental.
Their optimization is, however, a challenging issue. Even though affine
loop nests are generally present, the short trip counts and the
complexity of mathematical expressions make it hard to determine a
single or unique sequence of successful transformations. Therefore, we
present the design and systematic evaluation of COFFEE, a
domain-specific compiler for local assembly kernels. COFFEE manipulates
abstract syntax trees generated from a high-level domain-specific
language for PDEs by introducing domain-aware composable optimizations
aimed at improving instruction-level parallelism, especially SIMD
vectorization, and register locality. It then generates C code including
vector intrinsics. Experiments using a range of finite-element forms of
increasing complexity show that significant performance improvement is
achieved.
%
%
\end{abstract}





\acmformat{Fabio Luporini, Ana Lucia Varbanescu, Florian Rathgeber,
  Gheorghe-Teodor Bercea, J. Ramanujam, David A. Ham, and Paul
  H. J. Kelly, 2014. COFFEE: an Optimizing Compilerfor Finite Element
  Local Assembly.}


\begin{bottomstuff}

This research is partly funded by the MAPDES project, by the
Department of Computing at Imperial College London, by EPSRC through
grants EP/I00677X/1, EP/I006761/1, and EP/L000407/1, by NERC grants
NE/K008951/1 and NE/K006789/1, by the U.S.  National Science
Foundation through grants 0811457 and 0926687, by the U.S. Army
through contract W911NF-10-1-000, and by a HiPEAC collaboration
grant. The authors would like to thank Dr. Carlo Bertolli,
Dr. Lawrence Mitchell, and Dr. Francis Russell for their invaluable
suggestions and their contribution to the Firedrake project.

Author's addresses: Fabio Luporini $\&$ Florian Rathgeber $\&$
Gheorghe-Teodor Bercea $\&$ Paul H. J. Kelly, Department of Computing,
Imperial College London; Ana Lucia Varbanescu, Informatics Institute,
University of Amsterdam; David A. Ham, Department of Computing and
Department of Mathematics, Imperial College London; J. Ramanujam,
iCenter for Computation and Technology and the School of Elec.~Engr.\&
Comp.~Sci., Louisiana State University.
\end{bottomstuff}

\maketitle


\section{Introduction}

In many fields such as computational fluid dynamics, computational
electromagnetics and structural mechanics, phenomena are modelled by
partial differential equations (PDEs). Numerical techniques, like the
finite volume method and the finite element method, are widely
employed to approximate solutions of these PDEs. Unstructured meshes
are often used to discretize the computational domain, since they
allow an accurate representation of complex geometries. The
solution is sought by applying suitable numerical operations, or
kernels, to the entities of a mesh, such as edges, vertices, or
cells. On standard clusters of multicores, typically, a kernel is
executed sequentially by a thread, while parallelism is achieved by
partitioning the mesh and assigning each partition to a different node
or thread. Such an execution model, with minor variations, is adopted,
for example, in \cite{pyop2isc}, \cite{Fenics},
\cite{fluidity_manual_v4}, \cite{lizst}.



The time required to apply the numerical kernels is a major issue,
since the equation domain needs to be discretized into an extremely
large number of cells to obtain a satisfactory approximation
of the PDE, possibly of the order of trillions
\cite{Rossinelli2013}. For example, it has been well established that mesh
resolution is critical in the accuracy of numerical weather
forecasts. However, operational forecast centers have a strict time
limit in which to produce a forecast - 60 minutes in the case of the
UK Met Office. Producing efficient kernels has a direct scientific
payoff in higher resolution, and therefore more accurate,
forecasts. Computational cost is a dominant problem in
computational science simulations, especially for those based on
finite elements, which are the subject of this paper. We address, in
particular, the well-known problem of optimizing the local assembly
phase of the finite element method
\cite{Francis,quadrature1,petsc-integration-gpu,Kirby-FEM-opt}, which
can be responsible for a significant fraction of the overall computation
run-time, often in the range 30-60$\%$. With respect to these studies,
we propose a novel set of composable code transformations targeting, for
the first time, instruction-level parallelism - with emphasis on SIMD
vectorization - and register locality.

During the assembly phase, the solution of the PDE is approximated by
executing a problem-specific kernel over all cells, or elements, in the
discretized domain. In this work, we focus on relatively low order
finite element methods, in which an assembly kernel's working set is
usually small enough to fit the L1 cache. Low order methods are by no
means exotic: they are employed in a wide variety of fields, including
climate and ocean modeling, computational fluid dynamics, and structural
mechanics. High order methods such as the spectral element method
\cite{SPENCER} commonly require different algorithms to solve PDEs and
therefore are excluded from our study.

An assembly kernel is characterized by the presence of an affine,
often non-perfect loop nest, in which individual loops are rather
small: their trip count rarely exceeds 30, and may be as low as 3 for
low order methods. In the innermost loop, a problem-specific,
compute intensive expression evaluates a two dimensional array,
representing the result of local assembly in an element of the
discretized domain. With such a kernel structure, we focus on aspects
like minimization of floating-point operations, register allocation
and instruction-level parallelism, especially in the form of SIMD
vectorization.

Achieving high-performance is non-trivial. The complexity of the
mathematical expressions, often characterized by a large number of
operations on constants and small matrices, makes it hard to determine
a single or specific sequence of transformations that is successfully
applicable to all problems. Loop trip counts are typically small and can
vary significantly, which further exacerbates the issue. A
compiler-based approach is, therefore, the only reasonable option to
obtain close-to-peak performance in a wide range of different local
assembly kernels. Optimizations like padding, generalized
loop-invariant code motion, vector-register tiling, and expression
splitting, as well as their composition, are essential. but none are
supported by state-of-the-art polyhedral and vendor compilers. BLAS
routines could theoretically be employed, although a fairly
complicated control and data flow analysis would be required to
automate identification and extraction of matrix-matrix multiplies. In
addition, as detailed in Section~\ref{sec:perf-eval-blas}, the small
dimension of the matrices involved and the potential loss in data
locality can limit or eliminate the performance gain of this approach.

In order to overcome the constraints of the available compilers and
specialized linear algebra libraries, we have automated a set of generic
and model-driven code transformations in COFFEE\footnote{COFFEE stands
for COmpiler For FinitE Element local assembly.}, a compiler for
optimizing local assembly kernels. COFFEE is integrated with
Firedrake~\cite{firedrake-code}, a system for solving PDEs through the
finite element method based on the PyOP2
abstraction~\cite{pyop2ws,pyop2isc}. All problems expressible with this
framework are supported by COFFEE, including equations that can be found
at the core of real-world simulations, such as those used in our
performance evaluation. In evaluating our code transformations for a
range of relevant problems, we vary two key parameters that impact
solution accuracy and kernel cost: the polynomial order of the method
(we investigate from $p=1$ to $p=4$) and the geometry of elements in the
discretized domain (2D triangle, 3D tetrahedron, 3D prism). From the
point of view of the generated code, these two parameters directly
impact the size of both loop nests and mathematical expressions, as
elaborated in Section~\ref{sec:real-examples}.


Our experiments show that the original generated code for non-trivial
assembly kernels, despite following state-of-the-art techniques,
remains suboptimal in the context of modern multicore
architectures. Our domain-aware cost-model-driven sequence of code
transformations, aimed at improving SIMD vectorization and register
data locality, can result in performance improvements up to
4.4$\times$ over original kernels, and around 70$\%$ of the test cases
obtain a speed up greater than 2$\times$. The contributions of this
paper are
\begin{enumerate}
\item An optimization strategy for finite element local assembly that
  exploits domain knowledge and goes beyond the limits of both vendor
  and research compilers.
\item The design and implementation of a compiler that automates the
  proposed code transformations for any problems expressible in
  Firedrake.
\item A systematic analysis using a suite of examples of real-world
  importance that is evidence of significant performance improvements on
  two Intel architectures, a Sandy Bridge CPU and the Xeon Phi.
\end{enumerate}

The paper is organized as follows. In Section~\ref{sec:background} we
provide some background on local assembly, show code generated by
Firedrake and emphasize the critical computational
aspects. Section~\ref{sec:code-transf} describes the various code
transformations, highlighting when and how domain knowledge has been
exploited. The design and implementation of our compiler is discussed
in Section~\ref{sec:pyop2-compiler}. Section~\ref{sec:perf-results}
shows performance results. Related work is described in
Section~\ref{sec:related-work}, while Section~\ref{sec:conclusions}
reviews our contributions in the light of our results, and identifies
priorities for future work.



\section{Background and Motivating Examples}
\label{sec:background}


\begin{figure}
\begin{alltt}
\footnotesize
\textbf{Input:} element matrix (2D array, initialized to
0), element coordinates (array), coefficient fields (array,
e.g. velocity)
\textbf{Output:} element matrix (2D array)
- Compute Jacobian from coordinates
- Define basis functions
- Compute element matrix in an affine loop nest
\end{alltt}
\caption{Structure of a local assembly kernel}
\label{code:general-structure}
\end{figure}

\begin{table}[b]
\tbl{Type and variable names used in the various listings to identify local assembly objects.}{
\begin{tabulary}{1.0\columnwidth}{C|C|C}
\hline
Object name & Type & Variable name(s) \\\hline
Determinant of the Jacobian matrix & double & det  \\
Inverse of the Jacobian matrix & double & K1, K2, ... \\
Coordinates & double** & coords\\
Fields & double** & w \\
Numerical integration weights & double[] & W \\
Basis functions (and derivatives) & double[][] & X, Y, X1, ... \\
Element matrix & double[][] & A\\ \hline
\end{tabulary}
}
\label{table:map-name-letters}
\end{table}

Local assembly is the computation of contributions of a specific cell in
the discretized domain to the linear system which yields the PDE
solution. The process consists of numerically evaluating
problem-specific integrals to produce a matrix and a vector
\cite{quadrature1,fluidity_manual_v4}, whose sizes depend on the order
of the method. This operation is applied to all cells in the discretized
domain. In this work we focus on local matrices, or ``element
matrices'', which are more costly to compute than element vectors.

Given a finite element description of the input problem, expressed
through the domain-specific Unified Form Language (UFL)~\cite{ufl},
Firedrake employs the FEniCS form compiler (FFC)~\cite{FFC-Compiler} to
generate a C-code kernel implementing assembly using a
numerical quadrature. This kernel can be applied to any
element in the mesh, which follows from a mathematical property of the
finite element method. The evaluation of a local matrix can be
reduced to integration on a fixed ``reference'' element --- a special
element that does not belong to the domain --- after a suitable change
of coordinates. Firedrake triggers the compilation of an assembly
kernel using an available vendor compiler, and manages its (parallel)
execution over all elements in the mesh. As already explained, the
subject of this paper is to enhance this execution model by adding an
optimization stage prior to the generation of C code.


The structure of a local assembly kernel is shown in
Figure~\ref{code:general-structure}. The inputs are a zero-initialized
two dimensional array used to store the element matrix, the element's
coordinates in the discretized domain, and coefficient fields, for
instance indicating the values of velocity or pressure in the
element. The output is the evaluated element matrix. The kernel body
can be logically split into three parts:
\begin{enumerate}
  \item Calculation of the Jacobian matrix, its determinant and its
    inverse required for the aforementioned change of coordinates from
    the reference element to the one being computed.
  \item Definition of basis functions used, intuitively, to interpolate
    the contribution to the PDE solution over the element. The choice of
    basis functions is expressed in UFL directly by users. In the
    generated code, they are represented as global read-only two
    dimensional arrays (i.e., using \texttt{static const} in C) of
    double precision floats.
  \item Evaluation of the element matrix in an affine loop nest, in which
    the integration is performed.
\end{enumerate}
Table~\ref{table:map-name-letters} shows the variable names
we will use in the upcoming code snippets to refer to the various
kernel objects.


\begin{algorithm}[t]
\SetAlgorithmName{LISTING}{}
\footnotesize
\KwSty{void} helmholtz(double A[3][3], double **coords) $\lbrace$\\
~~// K, det = Compute Jacobian (coords) \\
~~\\
~~\KwSty{static const double} W[3] = $\lbrace$...$\rbrace$\\
~~\KwSty{static const double} X$\_$D10[3][3] = $\lbrace\lbrace$...$\rbrace\rbrace$\\
~~\KwSty{static const double} X$\_$D01[3][3] = $\lbrace\lbrace$...$\rbrace\rbrace$\\
~~\\
~~\KwSty{for} (\KwSty{int} i = 0; i$<$3; i++) \\
~~~~\KwSty{for} (\KwSty{int} j = 0; j$<$3; j++) \\
~~~~~~\KwSty{for} (\KwSty{int} k = 0; k$<$3; k++) \\
~~~~~~~~A[j][k] += ((Y[i][k]*Y[i][j]+\\
~~~~~~~~~~~+((K1*X$\_$D10[i][k]+K3*X$\_$D01[i][k])*(K1*X$\_$D10[i][j]+K3*X$\_$D01[i][j]))+\\
~~~~~~~~~~~+((K0*X$\_$D10[i][k]+K2*X$\_$D01[i][k])*(K0*X$\_$D10[i][j]+K2*X$\_$D01[i][j])))*\\
~~~~~~~~~~~*det*W[i]);\\
$\rbrace$
\caption{Local assembly code generated by Firedrake for a Helmholtz problem on a 2D triangular mesh using Lagrange $p=1$ elements.}
\label{code:helmholtz}
\end{algorithm}

\begin{algorithm}[t]
\SetAlgorithmName{LISTING}{}
\footnotesize
\KwSty{void} burgers(double A[12][12], double **coords, double **w) $\lbrace$\\
~~// K, det = Compute Jacobian (coords) \\
~~\\
~~\KwSty{static const double} W[5] = $\lbrace$...$\rbrace$\\
~~\KwSty{static const double} X1$\_$D001[5][12] = $\lbrace\lbrace$...$\rbrace\rbrace$\\
~~\KwSty{static const double} X2$\_$D001[5][12] = $\lbrace\lbrace$...$\rbrace\rbrace$\\
~~//11 other basis functions definitions.\\
~~...\\
~~\KwSty{for} (\KwSty{int} i = 0; i$<$5; i++) $\lbrace$\\
~~~~\KwSty{double} F0 = 0.0;\\
~~~~//10 other declarations (F1, F2,...)\\
~~~~...\\
~~~~\KwSty{for} (\KwSty{int} r = 0; r$<$12; r++) $\lbrace$\\
~~~~~~F0 += (w[r][0]*X1$\_$D100[i][r]);\\
~~~~~~//10 analogous statements (F1, F2, ...)\\
~~~~~~...\\
~~~~$\rbrace$\\
~~~~\KwSty{for} (\KwSty{int} j = 0; j$<$12; j++) \\
~~~~~~\KwSty{for} (\KwSty{int} k = 0; k$<$12; k++) \\
~~~~~~~~A[j][k] += (..(K5*F9)+(K8*F10))*Y1[i][j])+\\
~~~~~~~~~+(((K0*X1$\_$D100[i][k])+(K3*X1$\_$D010[i][k])+(K6*X1$\_$D001[i][k]))*Y2[i][j]))*F11)+\\
~~~~~~~~~+(..((K2*X2$\_$D100[i][k])+...+(K8*X2$\_$D001[i][k]))*((K2*X2$\_$D100[i][j])+...+(K8*X2$\_$D001[i][j]))..)+\\
~~~~~~~~~+ $<$roughly a hundred sum/muls go here$>$)..)*\\
~~~~~~~~~*det*W[i]);\\
~~$\rbrace$ \\
$\rbrace$
\caption{Local assembly code generated by Firedrake for a Burgers
problem on a 3D tetrahedral mesh using Lagrange $p=1$ elements.}
\label{code:burgers}
\end{algorithm}

The actual complexity of a local assembly kernel depends on the finite
element problem being solved. In simpler cases, the loop nest is
perfect, has short trip counts (in the range 3--15), and the computation
reduces to a summation of a few products involving basis functions. An
example is provided in Listing~\ref{code:helmholtz}, which shows the
assembly kernel for a Helmholtz problem using Lagrange basis functions
on 2D elements with polynomial order $p=1$. In other scenarios, for
instance when solving the Burgers equation, the number of arrays
involved in the computation of the element matrix can be much larger.
The assembly code is given in Listing~\ref{code:burgers} and contains 14
unique arrays that are accessed, where the same array can be referenced
multiple times within the same expression. This may also require the
evaluation of constants in outer loops (called $F$ in the code) to act
as scaling factors of arrays. Trip counts grow proportionally to the
order of the method and arrays may be block-sparse. In addition to a
larger number of operations, more complex cases like the Burgers
equation are characterized by high register pressure.

The Helmholtz and Burgers equations exemplify a large class of problems
and, as such, will constitute our benchmark problems, along with the
Diffusion equation. We carefully motivate this choice in
Section~\ref{sec:real-examples}.

Despite the infinite variety of assembly kernels which Firedrake can
generate, it is still possible to identify common domain-specific
traits that can be exploited for effective code transformations and
SIMD vectorization. These include: 1) memory accesses along the three
loop dimensions are always unit stride; 2) the \texttt{j} and \texttt{k}
loops are interchangeable, whereas interchanges involving the $i$ loop
require pre-computation of values (e.g. the $F$ values in Burgers) and
introduction of temporary arrays, as explained in
Section~\ref{sec:code-transf}; 3) depending on the problem being
solved, the \texttt{j} and \texttt{k} loops could iterate over the
same iteration space; 4) most of the sub-expressions on the right hand
side of the element matrix computation depend on just two loops
(either \texttt{i}-\texttt{j} or \texttt{i}-\texttt{k}). In
Section~\ref{sec:code-transf} we show how to exploit these
observations to define a set of systematic, composable optimizations.

\section{Code Transformations}
\label{sec:code-transf}
The code transformations presented in this section are applicable to
all finite element problems that can be formulated in Firedrake. As
already emphasized, the structure of mathematical expressions
evaluating the element matrix and the variation in loop trip counts,
although typically limited to the order of tens of iterations, render
the optimization process challenging. It is not always the same set of
optimizations that bring performance closest to the machine peak. For
example, the Burgers problem in Listing~\ref{code:burgers}, given the
large number of arrays accessed, suffers from high register pressure,
whereas the Helmholtz problem in Listing~\ref{code:helmholtz} does
not; this intuitively suggests that the two problems require a
different treatment, based on an in-depth analysis of both data and
iteration spaces. Furthermore, domain-knowledge enables
transformations that a general-purpose compiler could not apply,
making the optimization space even larger. In this context, our goal
is to understand the relationship between distinct code
transformations, their impact on local assembly kernels, and to what
extent their composability is effective in a class of problems and
architectures.

\subsection{Padding and Data Alignment}
The absence of stencils renders the element matrix computation easily
auto-vectorizable by a vendor compiler. Nevertheless,
auto-vectorization is not efficient if data are not aligned and if the
length of the innermost loop is not a multiple of the vector length
$\mbox{\texttt{VL}}$, especially when the loops are small as in local assembly.

Data alignment is enforced in two steps. Firstly, all arrays are
allocated to addresses that are multiples of $\mbox{\texttt{VL}}$. Then, two
dimensional arrays are padded by rounding the number of columns to the
nearest multiple of $\mbox{\texttt{VL}}$. For instance, assume the original size of a
basis function array is 3$\times$3 and $\mbox{\texttt{VL}}=4$ (e.g. AVX processor,
with 256 bit long vector registers and 64-bit double-precision
floats). In this case, a padded version of the array will have size
3$\times$4. The compiler is explicitly informed about data alignment
using a suitable pragma.

Padding of all two dimensional arrays involved in the evaluation of
the element matrix also allows to safely round the loop trip count to
the nearest multiple of $\mbox{\texttt{VL}}$. This avoids the introduction of a
remainder (scalar) loop from the compiler, which would render
vectorization less efficient. These extra iterations only write to the
padded region of the element matrix, and therefore have no side
effects on the final result.

\subsection{Generalized Loop-invariant Code Motion}
\label{sec:code-transf-licm}
From the inspection of the codes in Listings~\ref{code:helmholtz}
and~\ref{code:burgers}, it can be noticed that the computation of $A$
involves evaluating many sub-expressions which only depend on two
iteration variables. Since symbols in most of these sub-expressions
are read-only variables, there is ample space for loop-invariant code
motion. Vendor compilers apply this technique, although not in the
systematic way we need for our assembly kernels. We want to overcome
two deficiencies that both Intel and GNU compilers exhibit. First,
they only identify sub-expressions that are invariant with respect to
the innermost loop. This is an issue for sub-expressions depending on
\texttt{i}-\texttt{k}, which are not automatically lifted in the loop
order \texttt{ijk}. Second, the hoisted code is scalar and therefore not
subjected to auto-vectorization.

We work around these limitations with source-level loop-invariant code
motion. In particular, we pre-compute all values that an invariant
sub-expression assumes along its fastest varying dimension. This is
implemented by introducing a temporary array per invariant
sub-expression and by adding a new loop to the nest. At the price of
extra memory for storing temporaries, the gain is that lifted terms
can be auto-vectorized as part of an inner loop. Given the short trip
counts of our loops, it is important to achieve auto-vectorization of
hoisted terms in order to minimize the percentage of scalar
instructions, which could otherwise be significant. It is also worth
noting that, in some problems (e.g. Helmholtz), invariant
sub-expressions along \texttt{j} are identical to those along
\texttt{k}, and both loops iterate over the same iteration space, as
anticipated in Section~\ref{sec:background}. In these cases, we safely
avoid redundant pre-computation.

Listing~\ref{code:helmholtz-licm} shows the Helmholtz assembly code
after the application of loop-invariant code motion, padding, and data
alignment.

\begin{algorithm}[t]
\SetAlgorithmName{LISTING}{}
\footnotesize
\KwSty{void} helmholtz(double A[3][4], double **coords) $\lbrace$\\
~~\KwSty{$\#$define} ALIGN $\_\_$attribute$\_\_$((aligned(32))) \\
~~// K, det = Compute Jacobian (coords) \\
~~\\
~~\KwSty{static const double} W[3] ALIGN = $\lbrace$...$\rbrace$\\
~~\KwSty{static const double} X$\_$D10[3][4] ALIGN = $\lbrace\lbrace$...$\rbrace\rbrace$\\
~~\KwSty{static const double} X$\_$D01[3][4] ALIGN = $\lbrace\lbrace$...$\rbrace\rbrace$\\
~~\\
~~\KwSty{for} (\KwSty{int} i = 0; i$<$3; i++) $\lbrace$ \\
~~~~double LI$\_$0[4] ALIGN;\\
~~~~double LI$\_$1[4] ALIGN;\\
~~~~\KwSty{for} (\KwSty{int} r = 0; r$<$4; r++) $\lbrace$ \\
~~~~~~LI$\_$0[r] = ((K1*X$\_$D10[i][r])+(K3*X$\_$D01[i][r]));\\
~~~~~~LI$\_$1[r] = ((K0*X$\_$D10[i][r])+(K2*X$\_$D01[i][r]));\\
~~~~$\rbrace$\\
~~~~\KwSty{for} (\KwSty{int} j = 0; j$<$3; j++) \\
~~~~~~\KwSty{$\#$pragma vector aligned}\\
~~~~~~\KwSty{for} (\KwSty{int} k = 0; k$<$4; k++) \\
~~~~~~~~A[j][k] += (Y[i][k]*Y[i][j]+LI$\_$0[k]*LI$\_$0[j]+LI$\_$1[k]*LI$\_$1[j])*det*W[i]);\\
~~$\rbrace$\\
$\rbrace$
\caption{Local assembly code for the Helmholtz problem in
  Listing~\ref{code:helmholtz} after application of padding, data
  alignment, and \emph{licm}, for an AVX architecture. In this
  example, sub-expressions invariant to \texttt{j} are identical to
  those invariant to \texttt{k}, so they can be precomputed once in
  the $r$ loop.}
\label{code:helmholtz-licm}
\end{algorithm}

\subsection{Model-driven Vector-register Tiling}
\begin{figure}[h]
\centerline{\includegraphics[scale=0.8]{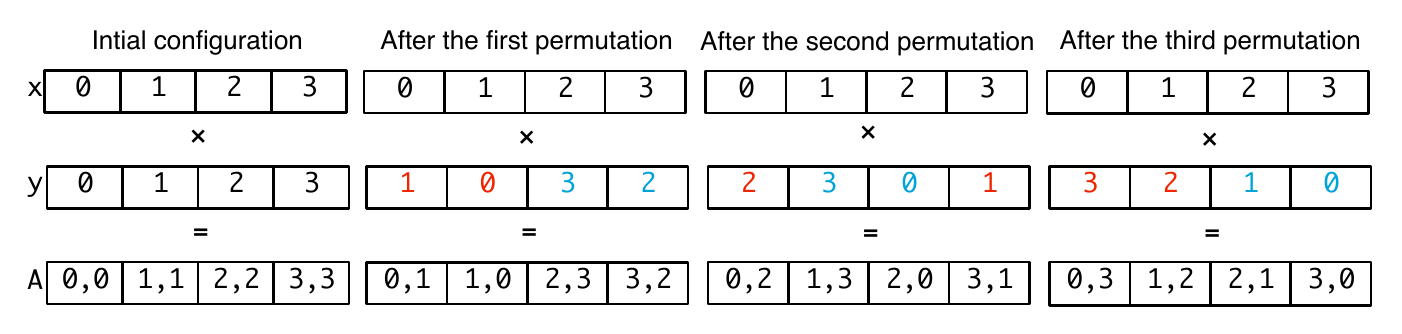}}
\caption{Outer-product vectorization by permuting values in a vector
  register.}
\label{fig:vect-by-vect}
\end{figure}

One notable problem of assembly kernels concerns register allocation
and register locality. The critical situation occurs when loop trip
counts and the variables accessed are such that the vector-register
pressure is high. Since the kernel's working set fits the L1 cache, it
is particularly important to optimize register management. Standard
optimizations, such as loop interchange, unroll, and unroll-and-jam,
can be employed to deal with this problem. In COFFEE, these
optimizations are supported either by means of explicit code
transformations (interchange, unroll-and-jam) or indirectly by
delegation to the compiler through standard pragmas (unroll). Tiling
at the level of vector registers is an additional feature of
COFFEE. Based on the observation that the evaluation of the element
matrix can be reduced to a summation of outer products along the
\texttt{j} and \texttt{k} dimensions, a model-driven vector-register
tiling strategy can be implemented. If we consider the code snippet in
Listing~\ref{code:helmholtz-licm} and we ignore the presence of the
operation \texttt{det*W3[i]}, the computation of the element matrix is
abstractly expressible as
\begin{equation}
\label{outer-product}
A_{jk} = \sum_{\substack{
  x \in B' \subseteq B \\
  y \in B'' \subseteq B}}
x_j\cdot y_k ~~~~~~ j,k = 0,...,2
\end{equation}
where $B$ is the set of all basis functions (or temporary variables,
e.g., \texttt{LI$\_$0}) accessed in the kernel, whereas $B'$ and $B''$
are generic problem-dependent subsets. Regardless of the specific
input problem, by abstracting from the presence of all variables
independent of both \texttt{j} and \texttt{k}, the element matrix
computation is always reducible to this form.
Figure~\ref{fig:vect-by-vect} illustrates how we can evaluate 16 entries
($j,k=0,...,3$) of the element matrix using just 2 vector registers,
which represent a 4$\times$4 tile, assuming $\vert B' \vert = \vert B''
\vert = 1$. Values in a register are shuffled each time a product is
performed. Standard compiler auto-vectorization for both GNU and Intel
compilers, instead, executes 4 broadcast operations (i.e., ``splat'' of
a value over all of the register locations) along the outer dimension to
perform the calculation. In addition to incurring a larger number of
cache accesses, it needs to keep between $f=1$ and $f=3$ extra registers
to perform the same 16 evaluations when unroll-and-jam is used, with $f$
being the unroll-and-jam factor.


\begin{figure}[h]
\centerline{\includegraphics[scale=0.8]{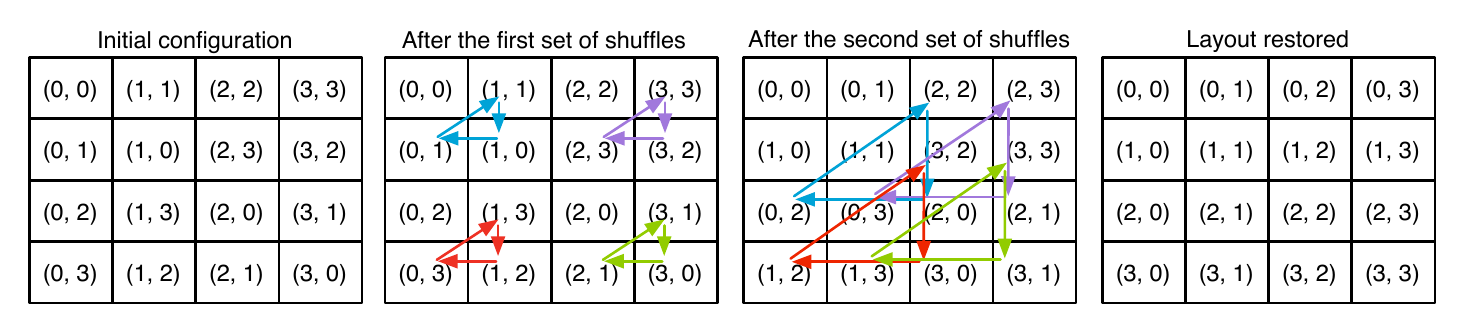}}
\caption{Restoring the storage layout after \emph{op-vect}. The figure
  shows how 4$\times$4 elements in the top-left block of the element
  matrix $A$ can be moved to their correct positions. Each rotation,
  represented by a group of three same-colored arrows, is implemented
  by a single shuffle intrinsic.}
\label{fig:restore-layout}
\end{figure}

The storage layout of $A$, however, is incorrect after the application
of this outer-product-based vectorization (\emph{op-vect}, in the
following). It can be efficiently restored with a sequence of vector
shuffles following the pattern highlighted in
Figure~\ref{fig:restore-layout}, executed once outside of the
\texttt{ijk} loop nest. The generated pseudo-code for the simple
Helmholtz problem when using \emph{op-vect} is shown in
Figure~\ref{code:helmholtz-opvect}.

\begin{algorithm}[t]
\SetAlgorithmName{LISTING}{}
\footnotesize
\KwSty{void} helmholtz(double A[4][4], double **coords) $\lbrace$\\
~~// K, det = Compute Jacobian (coords) \\
~~// Declaration of basis function matrices \\
~~\\
~~\KwSty{for} (\KwSty{int} i = 0; i$<$3; i++) $\lbrace$ \\
~~~~// Do generalized loop-invariant code motion \\
~~~~\KwSty{for} (\KwSty{int} j = 0; j$<$4; j+=4) \\
~~~~~~\KwSty{for} (\KwSty{int} k = 0; k$<$4; k+=4) $\lbrace$\\
~~~~~~~~// $load$ and $set$ intrinsics \\
~~~~~~~~// Compute A[0,0], A[1,1], A[2,2], A[3,3] \\
~~~~~~~~// One $permute\_pd$ intrinsic per \texttt{k}-loop $load$\\
~~~~~~~~// Compute A[0,1], A[1,0], A[2,3], A[3,2] \\
~~~~~~~~// One $permute2f128\_pd$ intrinsic per \texttt{k}-loop $load$\\
~~~~~~~~// ...\\
~~~~~~$\rbrace$\\
~~~~// Remainder loop (from $j=4$ to $j=6$)\\
~~$\rbrace$\\
~~// Restore the storage layout:\\
~~\KwSty{for} (\KwSty{int} j = 0; j$<$4; j+=4) $\lbrace$\\
~~~~$\_\_$m256d r0, r1, r2, r3, r4, r5, r6, r7;\\
~~~~\KwSty{for} (\KwSty{int} k = 0; k$<$4; k+=4) $\lbrace$\\
~~~~~~r0 = $\_$mm256$\_$load$\_$pd ($\&$A[j+0][k]);\\
~~~~~~// Load A[j+1][k], A[j+2][k], A[j+3][k]\\
~~~~~~r4 = $\_$mm256$\_$unpackhi$\_$pd (r1, r0);\\
~~~~~~r5 = $\_$mm256$\_$unpacklo$\_$pd (r0, r1);\\
~~~~~~r6 = $\_$mm256$\_$unpackhi$\_$pd (r2, r3);\\
~~~~~~r7 = $\_$mm256$\_$unpacklo$\_$pd (r3, r2);\\
~~~~~~r0 = $\_$mm256$\_$permute2f128$\_$pd (r5, r7, 32);\\
~~~~~~r1 = $\_$mm256$\_$permute2f128$\_$pd (r4, r6, 32);\\
~~~~~~r2 = $\_$mm256$\_$permute2f128$\_$pd (r7, r5, 49);\\
~~~~~~r3 = $\_$mm256$\_$permute2f128$\_$pd (r6, r4, 49);\\
~~~~~~$\_$mm256$\_$store$\_$pd ($\&$A[j+0][k], r0);\\
~~~~~~// Store A[j+1][k], A[j+2][k], A[j+3][k]\\
~~~~$\rbrace$\\
~~$\rbrace$\\
$\rbrace$
\caption{Local assembly code generated by Firedrake for the Helmholtz
  problem after application of \emph{op-vect} on top of the
  optimizations shown in Listing~\ref{code:helmholtz-licm}. In this
  example, the unroll-and-jam factor is 1.}
\label{code:helmholtz-opvect}
\end{algorithm}

\subsection{Expression Splitting}
\label{sec:expr-split}

\begin{algorithm}
\SetAlgorithmName{LISTING}{}
\footnotesize
\KwSty{void} helmholtz(double A[3][4], double **coords) $\lbrace$\\
~~~~// Same code as in Listing~\ref{code:helmholtz-licm} up to the j loop\\
~~~~\KwSty{for} (\KwSty{int} j = 0; j$<$3; j++) \\
~~~~~~\KwSty{$\#$pragma vector aligned}\\
~~~~~~\KwSty{for} (\KwSty{int} k = 0; k$<$4; k++) \\
~~~~~~~~A[j][k] += (Y[i][k]*Y[i][j]+LI$\_$0[k]*LI$\_$0[j])*det*W3[i];\\
~~~~\KwSty{for} (\KwSty{int} j = 0; j$<$3; j++) \\
~~~~~~\KwSty{$\#$pragma vector aligned}\\
~~~~~~\KwSty{for} (\KwSty{int} k = 0; k$<$4; k++) \\
~~~~~~~~A[j][k] += LI$\_$1[k]*LI$\_$1[j]*det*W3[i];\\
$\rbrace$
\caption{Local assembly code generated by Firedrake for the Helmholtz
  problem in which \emph{split} has been applied on top of the
  optimizations shown in Listing~\ref{code:helmholtz-licm}. In this
  example, the split factor is 2.}
\label{code:helmholtz-split}
\end{algorithm}

In complex kernels, like Burgers in Listing~\ref{code:burgers}, and on
certain architectures, achieving effective register allocation can be
challenging. If the number of variables independent of the
innermost-loop dimension is close to or greater than the number of
available CPU registers, poor register reuse is likely. This usually
happens when the number of basis function arrays, temporaries introduced
by generalized loop-invariant code motion, and problem constants is
large. For example, applying loop-invariant code motion to Burgers on a
3D mesh requires 24 temporaries for the \texttt{ijk} loop order. This
can make hoisting of the invariant loads out of the \texttt{k} loop
inefficient on architectures with a relatively low number of registers.
One potential solution to this problem consists of suitably
``splitting'' the computation of the element matrix $A$ into multiple
sub-expressions; an example, for the Helmholtz problem, is given in
Listing~\ref{code:helmholtz-split}. The transformation can be regarded
as a special case of classic loop fission, in which associativity of the
sum is exploited to distribute the expression across multiple loops. To
the best of our knowledge, expression splitting is not supported by
available compilers.

Splitting an expression has, however, several drawbacks. Firstly, it
increases the number of accesses to $A$ proportionally to the ``split
factor'', which is the number of sub-expressions produced. Also,
depending on how the split is executed, it can lead to redundant
computation. For example, the number of times the product $det*W3[i]$ is
performed is proportional to the number of sub-expressions, as shown in
the code snippet. Further, it increases loop overhead, for example
through additional branch instructions. Finally, it might affect
register locality: for instance, the same array could be accessed in
different sub-expressions, requiring a proportional number of loads be
performed. This is not the case for the Helmholtz example. Nevertheless,
as shown in Section~\ref{sec:perf-results}, the performance gain from
improved register reuse along inner dimensions can still be greater,
especially if the split factor and the splitting itself use heuristics
to minimize the aforementioned issues.

\begin{table}[h]
\tbl{Overview of code transformations for Firedrake-generated assembly kernels.}{
\begin{tabulary}{1.0\columnwidth}{C|C}
\hline
Name (Abbreviation) & Parameter \\\hline
Generalized loop-invariant code motion (\emph{licm}) &   \\
Padding &  \\
Data Alignment & \\
Loop interchange      & loops  \\
Loop unrolling  & unroll factor \\
Outer-product vectorization (\emph{op-vect}) & tile size \\
Expression splitting (\emph{split}) & split point, split factor \\ \hline
\end{tabulary}
}
\label{table:code-transformations}
\end{table}

Table~\ref{table:code-transformations} summarizes the code
transformations described so far. Given that many of these
transformations depend on some parameters (e.g. tile size), we need a
mechanism to prune such a large optimization space. This aspect is
treated in Section~\ref{sec:pyop2-compiler}.

\section{Overview of COFFEE}
\label{sec:pyop2-compiler}

\begin{figure}[b]
\begin{center}
\includegraphics[scale=0.75]{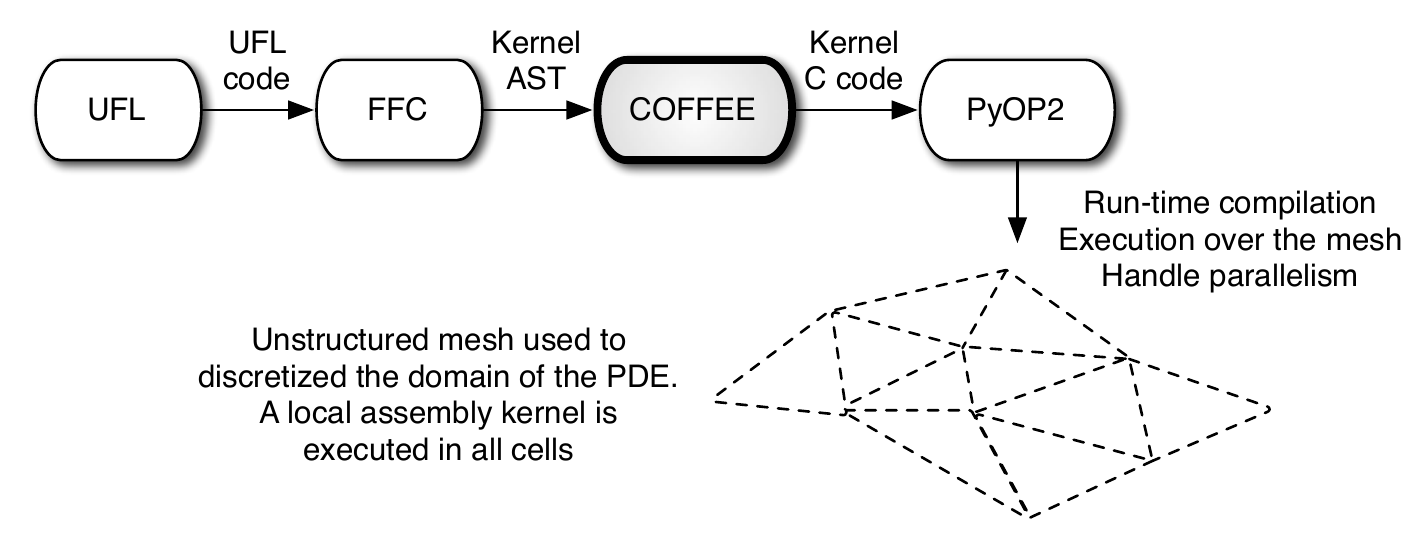}
\caption{High-level view of Firedrake. COFFEE is at the core,
  receiving ASTs from a modified version of the FEniCS Form compiler
  and producing optimized C code kernels.}
\label{fig:coffee-pipeline}
\end{center}
\end{figure}

Firedrake users employ the Unified Form Language to express problems in
a notation resembling mathematical equations. At run-time, the
high-level specification is translated by a modified version of the
FEniCS Form Compiler (FFC)~\cite{FFC-Compiler} into an abstract syntax
tree (AST) representation of one or more finite element assembly
kernels. ASTs are then passed to COFFEE to apply the transformations
described in Section~\ref{sec:code-transf}. The output of COFFEE, C
code, is eventually provided to PyOP2~\cite{pyop2ws,pyop2isc}, where
just-in-time compilation and execution over the discretized domain take
place. The flow is summarized in Figure~\ref{fig:coffee-pipeline}.

Because of the large number of parametric transformations, COFFEE
needs a mechanism to select the most suitable optimization strategy
for a given problem. Auto-tuning might be used, although it would
significantly increase the run-time overhead, since the generation of
ASTs occurs at run-time as soon as problem-specific data are
available. Our optimization strategy, based on heuristics and a simple
cost model, is described in the following, along with an overview of
the compiler.

The compiler structure is outlined in
Figure~\ref{algo:PyOP2Compiler}. Initially the AST is inspected,
looking for the presence of iteration spaces and domain-specific
information provided by the higher layer. If the kernel lacks an
iteration space, then so-called inter-kernel vectorization, in which
the outer, non-affine loop over mesh elements is vectorized, can be
applied. This feature, currently under development, has been shown to
be useful in several finite volume
applications~\cite{inter-kernel-vect}. Subsequently, an ordered
sequence of optimization steps are executed. Application of
\emph{licm} must precede padding and data alignment, due to the
introduction of temporary arrays. Based on a cost model, loop
interchange, \emph{split}, and \emph{op-vect} may be introduced. Their
implementation is based on analysis and transformation of the kernel's
AST. When \emph{op-vect} is selected, the compiler outputs AVX or
AVX-512 intrinsics code. Any possible corner cases are handled: for
example, if \emph{op-vect} is to be applied, but the size of the
iteration space is not a multiple of the vector length, then a
remainder loop, amenable to auto-vectorization, is inserted.

\begin{figure}[t]
\begin{alltt}
\footnotesize
\textbf{Input:}
ast: abstract syntax tree of a local assembly kernel produced by FFC
wrapper: the function within which the kernel is invoked
isa: represents the SIMD Instruction Set Architecture
\textbf{Output:} assembly code (C string)
\textbf{// Analyze ast and build optimization plan }
it\_space = analyze(ast)
\textbf{if not} it\_space
  ast.apply\_inter\_kernel\_vectorization(wrapper, isa)
  \textbf{return} wrapper + ast.from\_ast\_to\_c()
plan = cost\_model(it\_space.n\_inner\_arrays, isa.n\_regs)
\textbf{// Optimize ast based on plan}
ast.licm()
ast.padding()
ast.align\_data()
\textbf{if} plan.permute
  ast.permute\_assembly\_loops()
\textbf{if} plan.split\_factor
  ast.split(plan.split\_factor)
\textbf{if} plan.uaj\_factor
  uaj = MIN(plan.uaj\_factor, it\_space.j.size/isa.vf)
  ast.op\_vect(uaj)
\textbf{return} wrapper + ast.from\_ast\_to\_c()
\end{alltt}
\caption{Outline of the COFFEE architecture.}
\label{algo:PyOP2Compiler}
\end{figure}

All loops are interchangeable, provided that temporaries are
introduced if the nest is not perfect. For the employed storage
layout, the loop permutations \texttt{ijk} and \texttt{ikj} are likely
to maximize performance. Conceptually, this is motivated by the fact
that if the \texttt{i} loop were in an inner position, then a
significantly higher number of load instructions would be required at
every iteration. We tested this hypothesis in manually crafted
kernels. We found that the performance loss is greater than the gain
due to the possibility of accumulating increments in a register,
rather than memory, along the \texttt{i} loop. The choice between
\texttt{ijk} and \texttt{ikj} depends on the number of load
instructions that can be hoisted out of the innermost dimension. Our
compiler chooses as outermost the loop along which the number of
invariant loads is smaller so that more registers remain available to
carry out the computation of the element matrix.

Loop unroll (or unroll-and-jam of outer loops) is fundamental to the
exposure of instruction-level parallelism, and tuning the unroll
factor is particularly important. However, we noticed, by comparison
with implementations having manually-unrolled loops, that recent
versions of the Intel compiler estimate close-to-optimal unroll
factors when the loops are affine and their bounds are small and known
at compile-time, which is the case of our kernels. We therefore leave
the backend compiler in charge of selecting unroll factors. This
choice also simplifies COFFEE's cost model. The only situation in
which we explicitly unroll-and-jam a loop is when \emph{op-vect} is
used, since the transformed code prevents the Intel compiler from
applying this optimization, even if specific pragmas are added.


\begin{figure}
\begin{alltt}\footnotesize\internallinenumbers
\textbf{Input:} n\_outer\_arrays, n\_inner\_arrays, n\_consts, n\_regs
\textbf{Output:} uaj\_factor, split\_factor
n\_outer\_regs = n\_regs / 2
split\_factor = 0
\textbf{// Compute spltting factor }
\textbf{while} n\_outer\_arrays > n\_outer\_regs
  n\_outer\_arrays = n\_outer\_arrays / 2
  split\_factor = split\_factor + 1
\textbf{// Compute unroll-and-jam factor for \emph{op-vect}}
n\_regs\_avail = n\_regs - (n\_outer\_arrays + n\_consts)
uaj\_factor = n\_reg\_avail / n\_inner\_arrays
\textbf{// Estimate the benefit of permuting loops}
permute = n\_outer\_arrays > n\_inner\_arrays
\textbf{return} <permute, split\_factor, uaj\_factor>
\end{alltt}
\caption{The cost model is employed by the compiler to estimate the
  most suitable unroll-and-jam (when \emph{op-vect} is used) and split
  factors, avoiding the overhead of auto-tuning.}
\label{algo:applyCostModel}
\end{figure}

The cost model is shown in Figure~\ref{algo:applyCostModel}. It takes
into account the number of available logical vector registers,
\texttt{n\_regs}, and the number of unique variables accessed:
\texttt{n\_consts} counts variables independent of both \texttt{j} and
\texttt{k} loops and temporary registers, \texttt{n\_outer\_arrays}
counts \texttt{j}-dependent variables, and \texttt{n\_inner\_arrays}
counts \texttt{k}-dependent variables, assuming the \texttt{ijk} loop
order. These values are used to estimate unroll-and-jam and split
factors for \emph{op-vect} and \emph{split}. If a factor is 0, then
the corresponding transformation is not applied. The \emph{split}
transformation is triggered whenever the number of hoistable terms is
larger than the available registers along the outer dimension (lines
3-8), which is approximated as half of the total (line 3). A split
factor of $n$ means that the assembly expression should be ``cut''
into $n$ sub-expressions. Depending on the structure of the assembly
expression, each sub-expression might end up accessing a different
number of arrays; the cost model is simplified by assuming that all
sub-expressions are of the same size. The unroll-and-jam factor for
the \emph{op-vect} transformation is determined as a function of the
available logical registers, i.e., those not used for storing hoisted
terms (line 9-11). Finally, the profitability of loop interchange is
evaluated (line 13).

\section{Performance Evaluation}
\label{sec:perf-results}

\subsection{Experimental Setup}

Experiments were run on a single core of two Intel architectures, a
Sandy Bridge (I7-2600 CPU, running at 3.4GHz, 32KB L1 cache and 256KB
L2 cache) and a Xeon Phi (5110P, running at 1.05Ghz in native mode,
32KB L1 cache and 512KB L2 cache). We have chosen these two
architectures because of the differences in the number of logical
registers and SIMD lanes, which can impact the effectiveness of the
optimization strategy. The \texttt{icc 13.1} compiler was used. On the
Sandy Bridge, the compilation flags used were \texttt{-O2} and
\texttt{-xAVX} for auto-vectorization. On the Xeon Phi, optimization
level \texttt{-O3} was used. Other optimization levels performed, in
general, slightly worse.

\subsection{The Helmholtz, Diffusion, and Burgers equations}
\label{sec:real-examples}
Our code transformations were evaluated in three real-world problems
based on the following PDEs
\begin{enumerate}
\item Helmholtz
\item Advection-Diffusion
\item Burgers
\end{enumerate}
The three chosen benchmarks are \emph{real-life kernels} and comprise
the core differential operators in some of the most frequently
encountered finite element problems in scientific computing. The
Helmholtz and Diffusion kernels are archetypal second order elliptic
operators. They are complete and unsimplified examples of the
operators occurring for diffusion and viscosity in fluids, and for
imposing pressure in compressible fluids. As such, they are both
extensively used in climate and ocean modeling. Very similar
operators, for which the same optimisations are expected to be equally
effective, apply to elasticity problems, which are at the base of
computational structural mechanics. The Burgers kernel is a typical
example of a first order hyperbolic conservation law, which occurs in
real applications whenever a quantity is transported by a fluid (the
momentum itself, in our case). We chose this particular kernel since
it applies to a vector-valued quantity, while the elliptic operators
apply to scalar quantities; this impacts the generated code, as
explained next. In this way, the operators we have selected are
characteristic of both the second and first order operators that
dominate in fluids and solids simulations, and a wide variety on local
assembly codes can be tested.

The benchmarks were written in UFL (code available at~\cite{ufl-code})
and executed over real unstructured meshes through Firedrake. The
Helmholtz code has already been shown in
Listing~\ref{code:helmholtz}. For Advection-Diffusion, the Diffusion
equation, which uses the same differential operators as Helmholtz, is
considered. In the Diffusion kernel code, the main differences with
respect to Helmholtz are the absence of the $Y$ array and the presence
of a few more constants for computing the element matrix. Burgers is a
non-linear problem employing differential operators different from
those of Helmholtz and relying on vector-valued quantities, which has
a major impact on the generated assembly code (see
Listing~\ref{code:burgers}), where a larger number of basis function
arrays ($X1$, $X2$, ...) and constants ($F0$, $F1$, ..., $K0$, $K1$,
...) are generated.

These problems were studied varying both the shape of mesh elements
and the polynomial order $p$ of the method, whereas the element
family, Lagrange, is fixed. As might be expected, the larger the
element shape and $p$, the larger the iteration space. Triangles,
tetrahedron, and prisms were tested as element shape. For instance, in
the case of Helmholtz with $p=1$, the size of the \texttt{j} and
\texttt{k} loops for the three element shapes is, respectively, $3$,
$4$, and $6$. Moving to bigger shapes has the effect of increasing the
number of basis function arrays, since, intuitively, the behavior of
the equation has now to be approximated also along a third axis. On
the other hand, the polynomial order affects only the problem size
(the three loops \texttt{i}, \texttt{j}, and \texttt{k}, and, as a
consequence, the size of $X$ and $Y$ arrays). A range of polynomial
orders from $p=1$ to $p=4$ were tested; higher polynomial orders are
excluded from the study because of current Firedrake limitations. In
all these cases, the size of the element matrix rarely exceeds
30$\times$30, with a peak of 105$\times$105 in Burgers with prisms and
$p=4$.

\subsection{Loop permutation}
In the following, only results for the loop order \texttt{ijk} are
shown. For the considerations exposed in
Section~\ref{sec:pyop2-compiler}, loop interchanges having an inner
loop along \texttt{i} caused slow downs; also, interchanging
\texttt{j} and \texttt{k} loops while keeping \texttt{i} as outermost
loop did not provide any benefits.

\subsection{Impact of Generalized Loop-invariant Code Motion}
\label{licm-impact}

\newcommand{\licmresultsnorms}{
\begin{tabulary}{1.0\textwidth}{cccccc|cccc}
\cline{3-10}
& & \multicolumn{4}{c}{\texttt{Sandy Bridge}} & \multicolumn{4}{c}{\texttt{Xeon Phi}} \\
\cline{1-10}
\texttt{problem} & \texttt{shape} & \texttt{p1} & \texttt{p2} & \texttt{p3} & \texttt{p4} & \texttt{p1} & \texttt{p2} & \texttt{p3} & \texttt{p4} \\
\texttt{Helmholtz} & \texttt{triangle} & 1.05 & 1.46 & 1.68 & 1.67 & 1,49 & 1,06 & 1,05 & 1,17 \\
\texttt{Helmholtz} & \texttt{tetrahedron} & 1.36 & 2.10 & 2.64 & 2.27 & 1,28 & 1,29 & 2,05 & 1,73 \\
\texttt{Helmholtz} & \texttt{prism} & 2.16 & 2.28 & 2.45 & 2.06 & 1,04 & 2,26 & 1,93 & 1,64 \\[0.1cm]
\texttt{Diffusion} & \texttt{triangle} & 1.09 & 1.68 & 1.97 & 1.64 & 1,07 & 1,06 & 1,18 & 1,16 \\
\texttt{Diffusion} & \texttt{tetrahedron} & 1.30 & 2.20 & 3.12 & 2.60 & 1,00 & 1,38 & 2,02 & 1,74\\
\texttt{Diffusion} & \texttt{prism} & 2.15 & 1.82 & 2.71 & 2.32 & 1,11 & 2,16 & 1,85 & 2,83\\[0.1cm]
\texttt{Burgers} & \texttt{triangle} & 1.53 & 1.81 & 2.68 & 2.46 & 1,21 & 1,42 & 2,34 & 2,97  \\
\texttt{Burgers} & \texttt{tetrahedron} & 1.61 & 2.24 & 1.69 & 1.59 & 1,01 & 2,55 & 0,98 & 1,21  \\
\texttt{Burgers} & \texttt{prism} & 2.11 & 2.20 & 1.66 & 1.32 & 1,39 & 1,56 & 1,18 & 1,04 \\
\cline{1-10}
\end{tabulary}
}

\begin{table*}[t]
\tbl{Performance improvement due to generalized loop-invariant code
  motion, for different element shapes (triangle, tetrahedron, prism)
  and polynomial orders ($p \in [1, 4]$), over the original
  non-optimized code, for the Helmholtz, Diffusion and Burgers
  problems.}{ \scriptsize \licmresultsnorms }
\label{table:perf-results-licm}
\end{table*}
Table~\ref{table:perf-results-licm} illustrates the performance
improvement obtained when \emph{licm} is applied. In general, the
speed ups are notable. The main reasons were anticipated in
Section~\ref{sec:code-transf-licm}: in the original code, 1)
sub-expressions invariant to outer loops are not automatically
hoisted, while 2) sub-expressions invariant to the innermost loop are
hoisted, but their execution is not auto-vectorized. These
observations come from inspection of assembly code generated by the
compiler.

The gain tends to grow with the computational cost of the kernels:
bigger loop nests (i.e., larger element shapes and polynomial orders)
usually benefit from the reduction in redundant computation, even
though extra memory for the temporary arrays is required. Some
discrepancies to this trend are due to a less effective
auto-vectorization. For instance, on the Sandy Bridge, the improvement
at $p=3$ is larger than that at $p=4$ because, in the latter case, the
size of the innermost loop is not a multiple of the vector length, and
a reminder scalar loop is introduced at compile time. Since the loop
nest is small, the cost of executing the extra scalar iterations can
have a significant impact. The reminder loop overhead is more
pronounced on the Xeon Phi, where the vector length is twice as big,
which leads to proportionally larger scalar reminder loops.

\subsection{Impact of Padding and Data Alignment}

\newcommand{\licmapresultsnorms}{
\begin{tabulary}{1.0\textwidth}{cccccc|cccc}
\cline{3-10}
& & \multicolumn{4}{c}{\texttt{Sandy Bridge}} & \multicolumn{4}{c}{\texttt{Xeon Phi}} \\
\cline{1-10}
\texttt{problem} & \texttt{shape} & \texttt{p1} & \texttt{p2} & \texttt{p3} & \texttt{p4} & \texttt{p1} & \texttt{p2} & \texttt{p3} & \texttt{p4} \\[0.1cm]
\texttt{Helmholtz} & \texttt{triangle} & 1.32 & 1.88 & 2.87 & 4.13 & 1,50 & 2,41 & 1,30 & 1,96 \\
\texttt{Helmholtz} & \texttt{tetrahedron} & 1.35 & 3.32 & 2.66 & 3.27 & 1,41 & 1,50 & 2,79 & 2,81 \\
\texttt{Helmholtz} & \texttt{prism} & 2.63 & 2.74 & 2.43 & 2.75 & 2,38 & 2,47 & 2,15 & 1,71\\[0.1cm]
\texttt{Diffusion} & \texttt{triangle} & 1.38 & 1.99 & 3.07 & 4.28 & 1,08 & 1,88 & 1,20 & 1,97\\
\texttt{Diffusion} & \texttt{tetrahedron} & 1.41 & 3.70 & 3.18 & 3.82 & 1,05 & 1,51 & 2,76 & 3,00\\
\texttt{Diffusion} & \texttt{prism} & 2.55 & 3.13 & 2.73 & 2.69 & 2,41 & 2,52 & 2,05 & 2,48\\[0.1cm]
\texttt{Burgers} & \texttt{triangle} & 1.56 & 2.28 & 2.61 & 2.77 & 2,84 & 2,26 & 3,96 & 4,27 \\
\texttt{Burgers} & \texttt{tetrahedron} & 1.61 & 2.10 & 1.60 & 1.78 & 1,48 & 3,83 & 1,55 & 1,29 \\
\texttt{Burgers} & \texttt{prism} & 2.19 & 2.32 & 1.64 & 1.42 & 2,18 & 2,82 & 1,24 & 1,25 \\
\cline{1-10}
\end{tabulary}
}

\begin{table*}[t]
\tbl{Performance improvement due to generalized loop-invariant code
  motion, data alignment, and padding, for different element shapes
  (triangle, tetrahedron, prism) and polynomial orders ($p \in [1,
    4]$), over the original non-optimized code, for the Helmholtz,
  Diffusion and Burgers problems.}{ \scriptsize \licmapresultsnorms }
\label{table:perf-results-licmap}
\end{table*}

Table~\ref{table:perf-results-licmap} shows the cumulative impact of
\emph{licm}, data alignment, and padding over the original code. In
the following, this version of the code is referred to as
\emph{licm-ap}. Padding, which avoids the introduction of a reminder
loop as described in Section~\ref{licm-impact}, as well as data
alignment, enhance the quality of auto-vectorization. Occasionally the
run-time of \emph{licm-ap} is close to that of \emph{licm}, since the
non-padded element matrix size is already a multiple of the vector
length. Rarely \emph{licm-ap} is slower than \emph{licm} (e.g. in
Burgers $p=3$ on the Sandy Bridge). One possible explanation is that
the number of aligned temporaries introduced by \emph{licm} is so
large to induce cache associativity conflicts.

\subsection{Impact of Vector-register Tiling}
\label{sec:perf-eval-opvect}
In this section, we evaluate the impact of vector-register tiling. We
compare two versions: the baseline, \emph{licm-ap}; and
vector-register tiling on top of \emph{licm-ap}, which in the
following is referred to simply as \emph{op-vect}.

\begin{figure}[h]
	\centering
	\subfloat[Sandy Bridge]{	\includegraphics[scale=0.52]{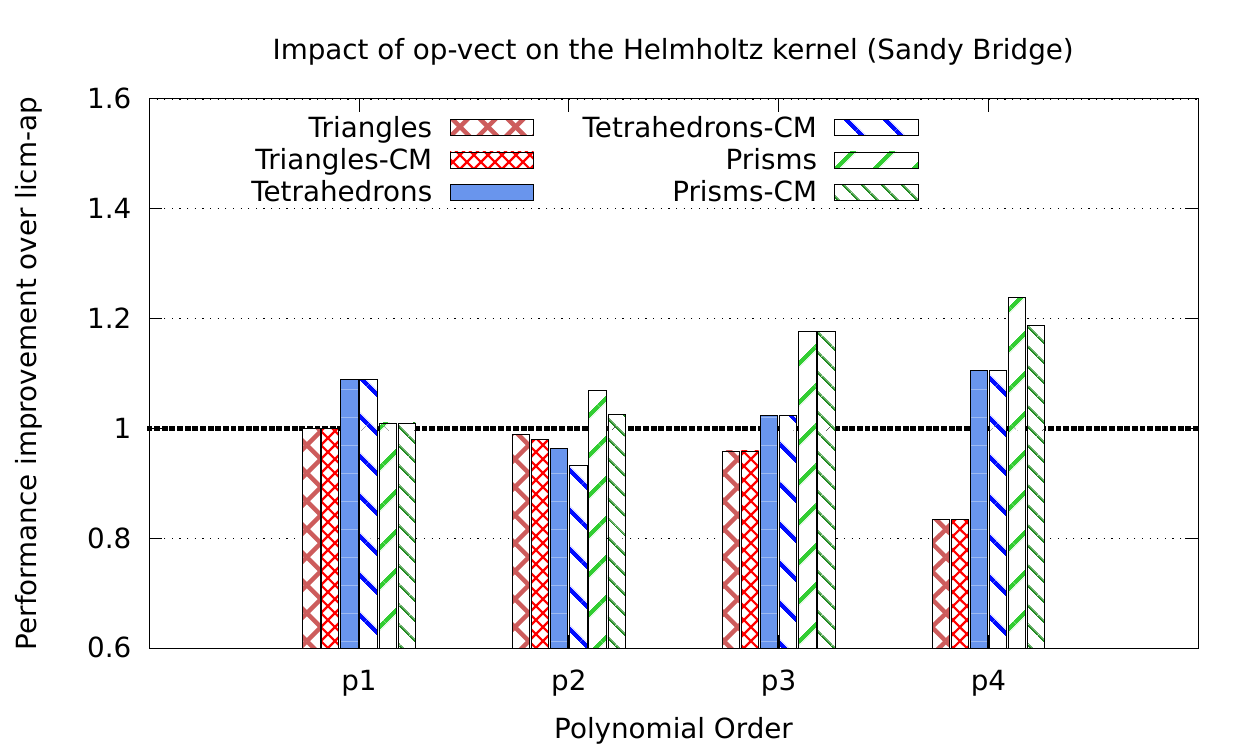}}
	\subfloat[Xeon Phi]{	\includegraphics[scale=0.52]{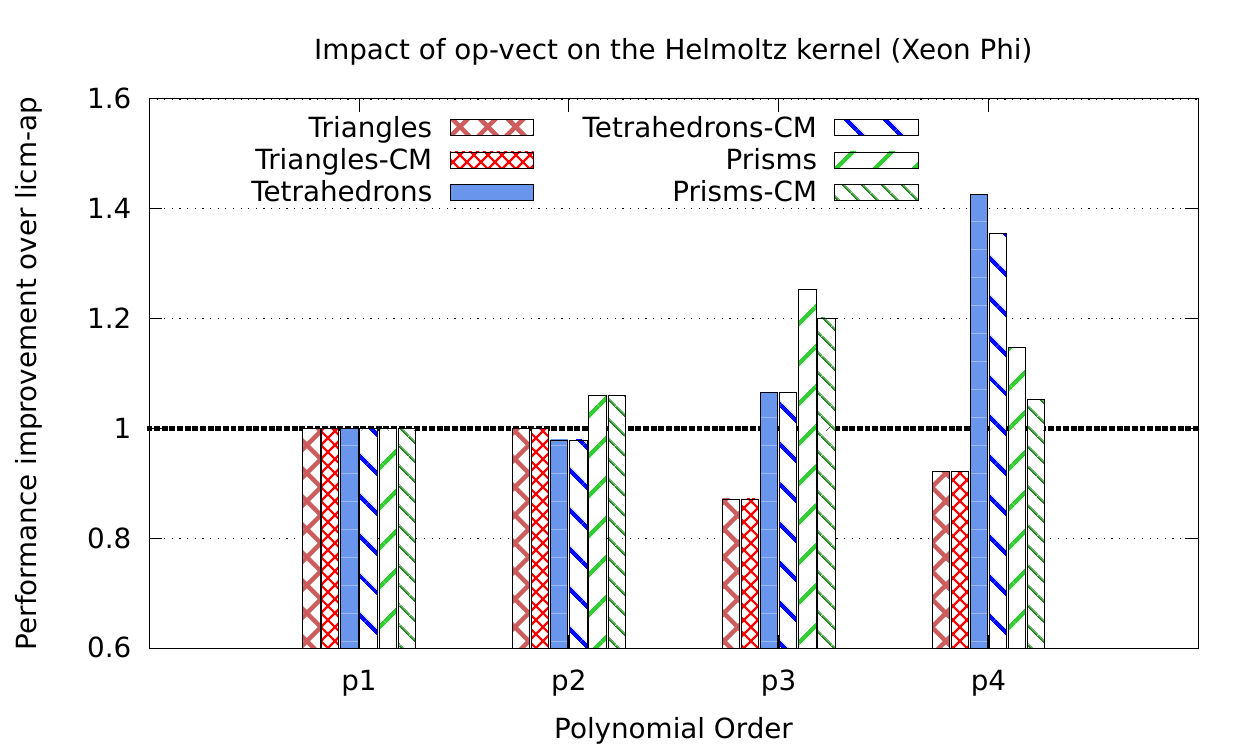}}
	\caption{Performance improvement over \emph{licm-ap} obtained
          by \emph{op-vect} in the Helmholtz kernel. Bars suffixed
          with ``CM'' indicate that the cost model was used to
          transform the kernel.}
	\label{fig:opvect-helmholtz-speedup}
\end{figure}

\begin{figure}[h]
	\centering
	\subfloat[Sandy Bridge]{	\includegraphics[scale=0.52]{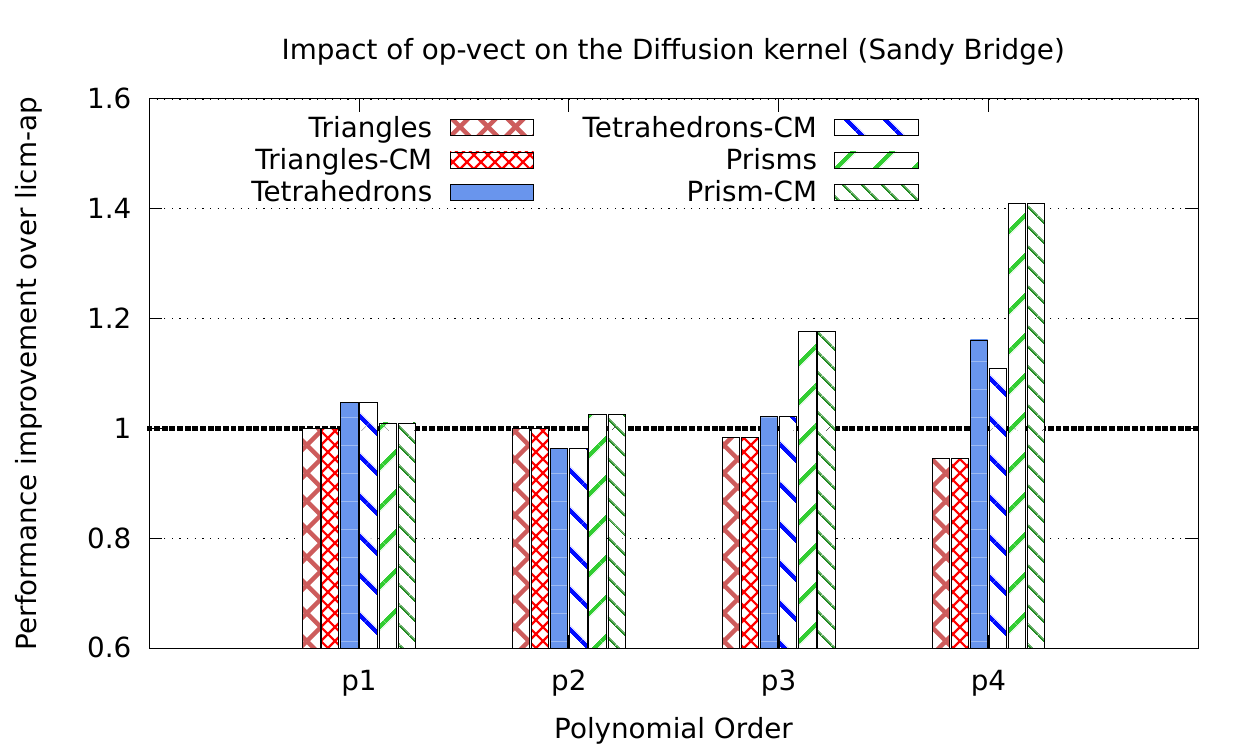}}
	\subfloat[Xeon Phi]{	\includegraphics[scale=0.52]{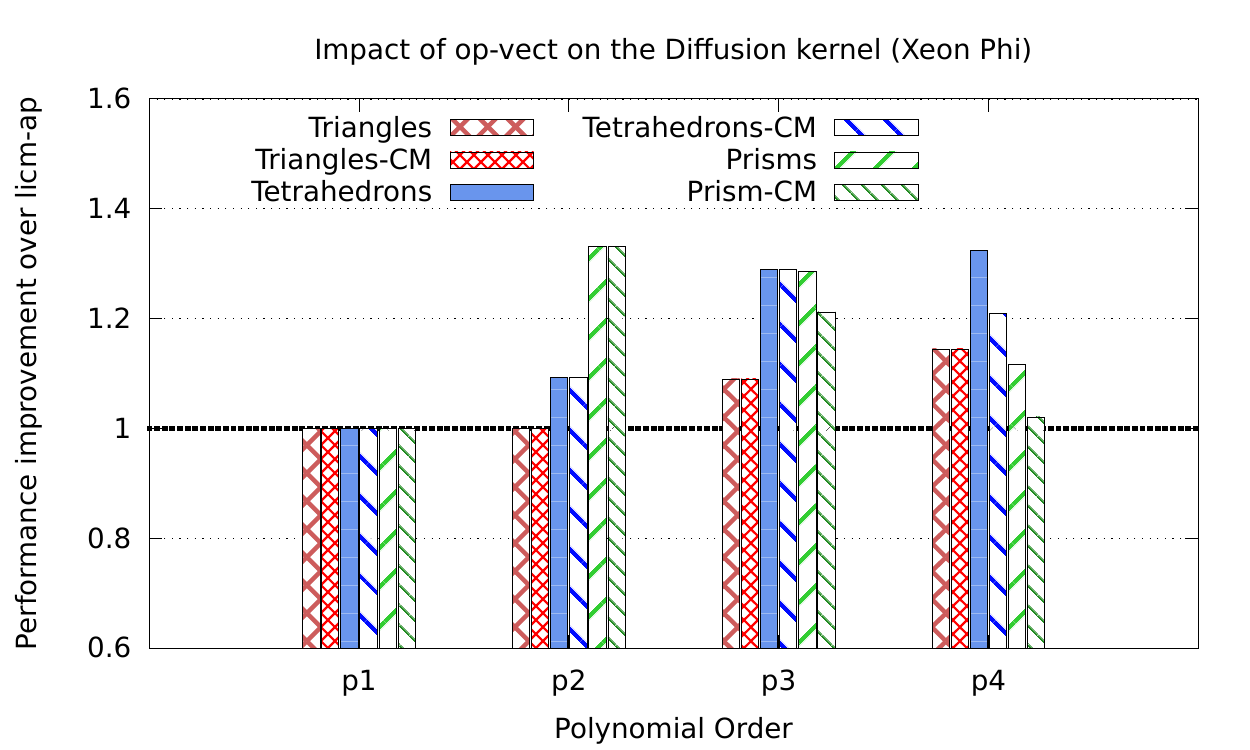}}
	\caption{Performance improvement over \emph{licm-ap} obtained
          by \emph{op-vect} in the Diffusion kernel. Bars suffixed
          with ``CM'' indicate that the cost model was used to
          transform the kernel.}
	\label{fig:opvect-diffusion-speedup}
\end{figure}

Figures~\ref{fig:opvect-helmholtz-speedup}
and~\ref{fig:opvect-diffusion-speedup} illustrate the speed up
achieved by \emph{op-vect} over \emph{licm-ap} in the Helmoltz and
Diffusion kernels, respectively. As explained in
Section~\ref{sec:pyop2-compiler}, the \emph{op-vect} version requires
the unroll-and-jam factor to be explicitly set. To distinguish between
the two ways this parameter was determined, for each problem instance
(equation, element shape, polynomial order) we report two bars: one
shows the best speed-up obtained after all feasible unroll-and-jam
factors were tried; the other shows the speed up when the
unroll-and-jam factor was retrieved via the cost model. In a plot
legend, cost model bars are suffixed with ``CM''.

It is worth noticing that, in most cases, the cost model successfully
determines how to transform a kernel to maximize its performance. This
is chiefly because assembly kernels fit the L1 cache, so, within a
certain degree of confidence, it is possible to predict how to obtain
a fast implementation by simply reasoning on the register
pressure. For each problem, the cost model stated to use the default
loop permutation, to apply a particular unroll-and-jam factor, and not
to perform expression splitting, which, as explained in
Section~\ref{sec:perf-results-split}, only deteriorates performance in
Helmholtz and Diffusion.

The rationale behind these results is that the effect of
\emph{op-vect} is significant in problems in which the assembly loop
nest is relatively big. When the loops are short, since the number of
arrays accessed at every loop iteration is rather small (between 4 and
8 temporaries, plus the element matrix itself), there is no need for
vector-register tiling; extensive unrolling is sufficient to improve
register re-use and, therefore, to maximize the performance. However,
as the iteration space becomes larger, \emph{op-vect} leads to
improvements up to 1.4$\times$ on the Sandy Bridge (Diffusion,
prismatic mesh, $p=4$ - increasing the overall speed up from
2.69$\times$ to 3.87$\times$), and up to 1.4$\times$ on the Xeon Phi
(Helmholtz, tetrahedral mesh, $p=3$ - bringing the overall speed up
from 1.71$\times$ to 2.42$\times$).

Using the Intel Architecture Code Analyzer tool~\cite{IACA} on the
Sandy Bridge, we confirmed that speed ups are a consequence of
increased register re-use. In Helmholtz $p=4$, for example, the tool
showed that when using \emph{op-vect} the number of clock cycles to
execute one iteration of the \texttt{j} loop decreases by roughly
17$\%$, and that this is a result of the relieved pressure on both of
the data (cache) ports available in the core.

On the Sandy Bridge, we have also measured the performance of
individual kernels in terms of floating-point operations per
second. The theoretical peak on a single core, with the Intel Turbo
Boost technology activated, is 30.4 GFlop/s. In the case of Diffusion
using a prismatic mesh and $p=4$, we achieved a maximum of 21.9
GFlop/s with \emph{op-vect} enabled, whereas 16.4 GFlop/s was obtained
when only \emph{licm-ap} is used. This result is in line with the
expectations: analysis of assembly code showed that, in the
\texttt{jk} loop nest, which in this problem represents the bulk of
the computation, 73$\%$ of instructions are actually floating-point
operations.

Application of \emph{op-vect} to the Burgers problem induces
significant slow downs due to the large number of temporary arrays
that need to be tiled, which exceeds the available logical registers
on the underlying architecture. Expression splitting can be used in
combination with \emph{op-vect} to alleviate this issue; this is
discussed in the next section.

\subsection{Impact of Expression Splitting}
\label{sec:perf-results-split}
Expression splitting relieves the register pressure when the element
matrix evaluation needs to read from a large number of basis function
arrays. As detailed in Section~\ref{sec:expr-split}, in the price to
pay for this optimazion, there are an increased number of accesses to
the element matrix and, potentially, redundant computation. Similarly
to the analysis of vector-register tiling, we compare two versions:
the baseline, \emph{licm-ap}; and expression splitting on top of
\emph{licm-ap}, which, for simplicity, in the following is referred to
as \emph{split}.

For the Helmholtz and Diffusion kernels, in which only between 4 and 8
temporaries are read at every loop iteration, \texttt{split} tends to
slow down the computation, because of the aforementioned
drawbacks. Slow downs up to 1.4$\times$ and up to 1.6$\times$ were
observed, respectively, on the Sandy Bridge and the Xeon Phi. Note
that the cost model prevents the adoption of the transformation: the
\texttt{while} statement in Figure~\ref{algo:applyCostModel} is indeed
never entered.

\begin{figure}[h]
	\centering
	\subfloat[Sandy Bridge]{	\includegraphics[scale=0.52]{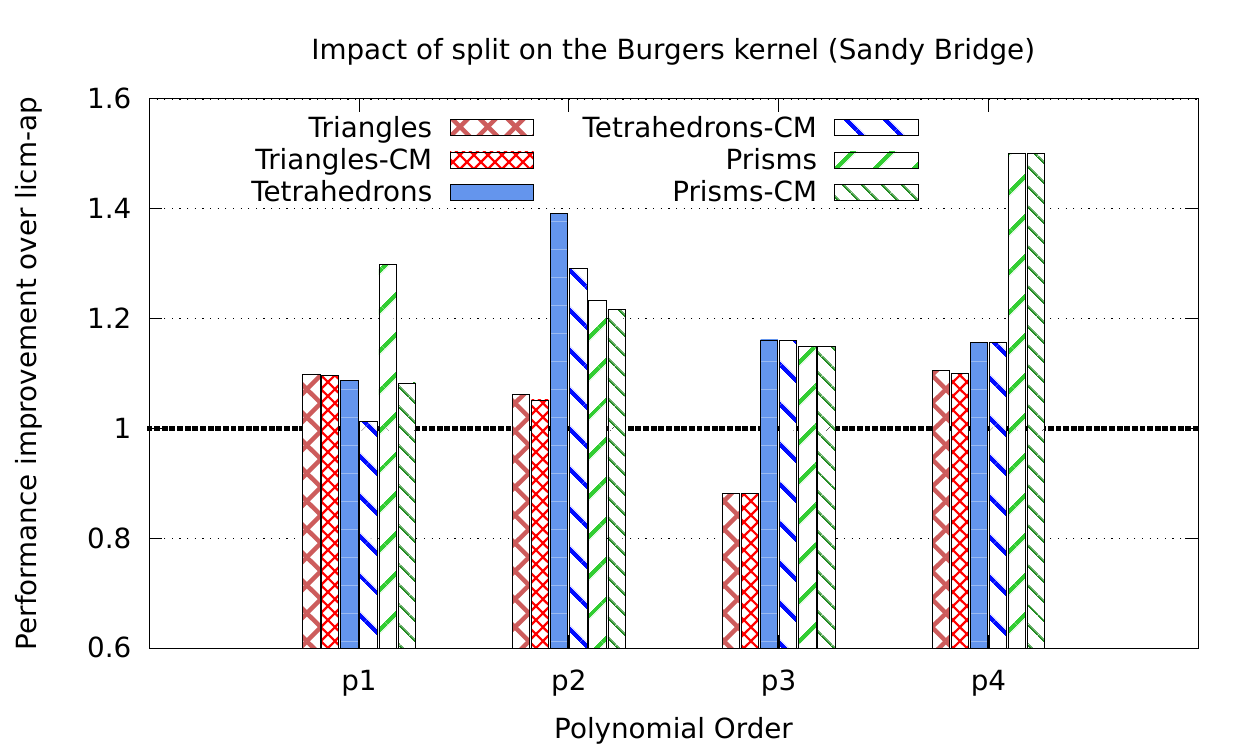}}
	\subfloat[Xeon Phi]{	\includegraphics[scale=0.52]{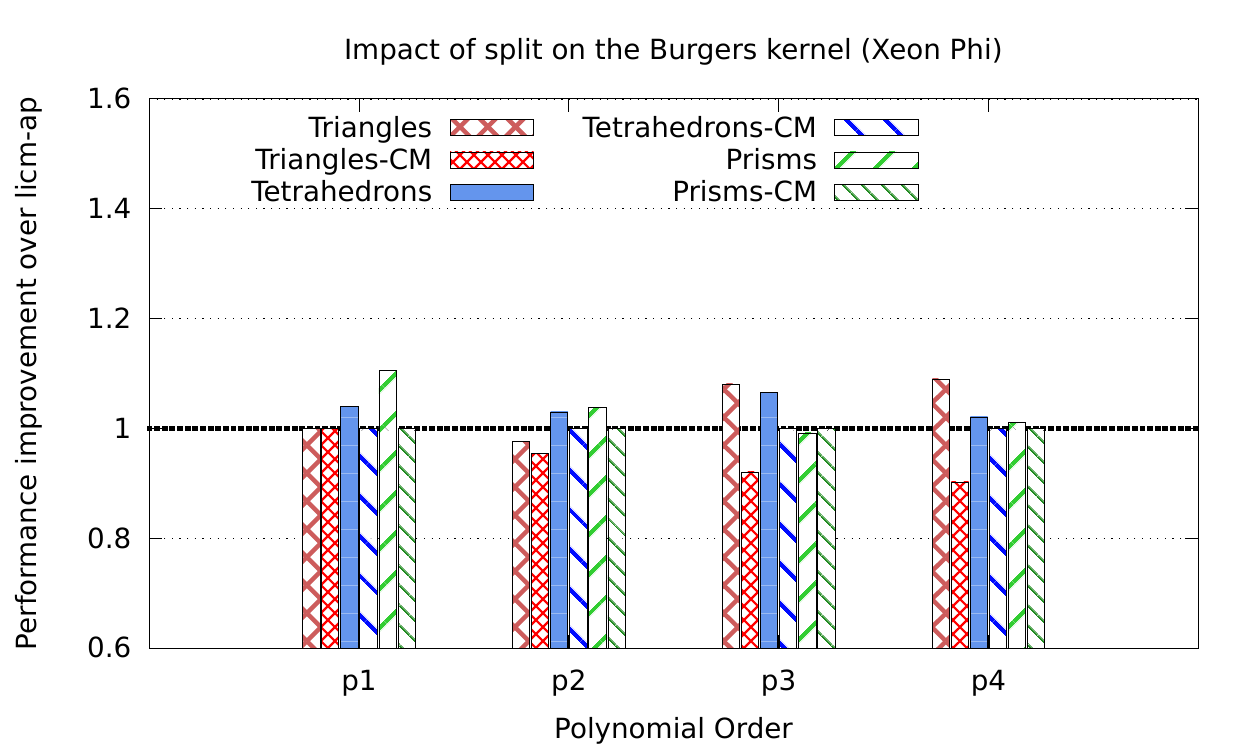}}
	\caption{Performance improvement over \emph{licm-ap} obtained
          by \emph{split} in the Burgers kernel. Bars suffixed with
          ``CM'' indicate that the cost model was used to transform
          the kernel.}
	\label{fig:split-burgers-speedup}
\end{figure}

In the Burgers kernels, between 12 and 24 temporaries are accessed at
every loop iteration, so \emph{split} plays a key role on the Sandy
Bridge, where the number of available logical registers is only
16. Figure~\ref{fig:split-burgers-speedup} shows the performance
improvement achieved by \emph{split} over \emph{licm-ap}. In almost
all cases, a split factor of 1, meaning that the original expression
was divided into two parts, ensured close-to-peak perforance. The
transformation negligibly affected register locality, so speed ups up
to 1.5$\times$ were observed. For instance, on the Sandy Bridge, when
$p=4$ and a prismatic mesh is employed, the overall performance
improvement (i.e., the one over the original code) increases from
1.44$\times$ to 2.11$\times$. On the Xeon Phi, the impact of
\emph{split} is only marginal, since register spilling is limited by
the presence of 32 logical vector units.

On the Sandy Bridge, the performance of the Burgers kernel on a
prismatic mesh was 20.0 GFlop/s from $p=1$ to $p=3$, while it was 21.3
GFlop/s in the case of $p=4$. These values are notably close to the
peak performance of 30.4 GFlop/s. Disabling \emph{split} makes the
performance drop to 17.0 GFlop/s for $p=1, 2$, 18.2 GFlop/s for $p=3$,
and 14.3 GFlop/s for $p=4$. These values are in line with the speedups
shown in Figure~\ref{fig:split-burgers-speedup}.

The \emph{split} transformation was also tried in combination with
\emph{op-vect} (\emph{split-op-vect}), although the cost model
prevents its adoption on both platforms. Despite improvements up to
1.22$\times$, \emph{split-op-vect} never outperforms
\emph{split}. This is motivated by two factors: for small split
factors, such as 1 and 2, the data space to be tiled is still too big,
and register spilling affects run-time; for higher ones,
sub-expressions become so small that, as explained in
Section~\ref{sec:perf-eval-opvect}, extensive unrolling already allows
to achieve a certain degree of register re-use.

\subsection{Comparison with FEniCS Form Compiler's built-in Optimizations}
\label{sec:perf-results-fenics}
We have modified the FEniCS Form Compiler (FFC) to return an abstract
syntax tree representation of a local assembly kernel, rather than
plain C++ code, so as to enable code transformations in
COFFEE. Besides Firedrake, FFC is used in the FEniCS
project~\cite{Fenics}. In FEniCS, FFC can apply its own model-driven
optimizations to local assembly kernels~\cite{quadrature1}, which
mainly consist of loop-invariant code motion and elimination, at code
generation time, of floating point operations involving zero-valued
entries in basis function arrays.

The FEniCS Form Compiler's loop-invariant code motion is different
from COFFEE's. It is based on expansion of arithmetic operations, for
example applying distributivity and associativity to products and sums
at code generation time, to identify terms that are invariant of the
whole loop nest. Depending on the way expansion is performed,
operation count may not decrease significantly.

Elimination of zero-valued terms, which are the result of using
vector-valued quantities in the finite element problem, has the effect
of introducing indirection arrays in the generated code. This kind of
optimization is currently under development in COFFEE, although it
will differ from FEniCS' by avoiding non-contiguous memory accesses,
which would otherwise affect vectorization, at the price of removing
fewer zero-valued contributions.

Table~\ref{table:comparison-to-FFC-opt} summarizes the performance
achieved by COFFEE over the \emph{fastest} FEniCS (FFC) implementation
on the Sandy Bridge for the Burgers, Helmholtz and Diffusion
kernels. Burgers' slow downs occur in presence of a small iteration
space (triangular mesh, $p \in [1, 2]$; tetrahedral mesh, $p \in [1,
  2]$; prismatic mesh, $p = 1$). The result shown represents the worst
slow down, which was obtained with a triangular mesh and $p = 1$. This
is a result of removing zero-valued entries in FEniCS' basis function
arrays: some operations are avoided, but indirection arrays prevent
auto-vectorization, which significantly impacts performance as soon as
the element matrix becomes bigger than 12$\times$12. However, with the
forthcoming zero-removal optimization in COFFEE, we expect to
outperform FEniCS in all problems. In the cases of Helmholtz and
Diffusion, the minimum improvements are, respectively, 1.10$\times$
and 1.18$\times$ (2D mesh, $p=1$), which tend to increase with
polynomial order and element shape up to the values illustrated in the
table.

\begin{table}[h]
\tbl{Performance comparison between FEniCS and COFFEE on the Sandy Bridge.}{
\begin{tabulary}{1.0\columnwidth}{C|C|C}
\hline
Problem & Max slow down & Max speed up \\\hline
Helmholtz & - & 4.14$\times$ \\
Diffusion & - & 4.28$\times$ \\
Burgers & 2.24$\times$ & 1.61$\times$ \\\hline
\end{tabulary}
}
\label{table:comparison-to-FFC-opt}
\end{table}

\subsection{Comparison with hand-made BLAS-based implementations}
\label{sec:perf-eval-blas}
For the Helmholtz problem on a tetrahedral mesh, manual
implementations based on \texttt{Intel MKL BLAS} were tested on the
Sandy Bridge. This particular kernel can be easily reduced to a
sequence of four matrix-matrix multiplies that can be computed via
calls to BLAS \texttt{dgemm}. In the case of $p=4$, where the element
matrix is of size 35$\times$35, the computation was almost twice
slower than the case in which \emph{licm-ap} was used, with the slow
down being even worse for smaller problem sizes. These experiments
suggest that the question regarding to what extent linear algebra
libraries can improve performance cannot be trivially answered. This
is due to a combination of issues: the potential loss in data
locality, as exposed in Section~\ref{sec:expr-split}, the actual
effectiveness of these libraries when the arrays are relatively small,
and the problem inherent to assembly kernels concerning extraction of
matrix-matrix multiplies from static analysis of the kernel's code. A
comprehensive study of these aspects will be addressed in further
work.

\section{Related Work}
\label{sec:related-work}
The finite element method is extensively used to approximate solutions
of PDEs. Well-known frameworks and applications include
nek5000~\cite{nek5000-web-page}, the FEniCS project~\cite{Fenics},
Fluidity~\cite{fluidity_manual_v4}, and of course Firedrake. Numerical
integration is usually employed to implement the local assembly
phase. The recent introduction of domain specific languages (DSLs) to
decouple the finite element specification from its underlying
implementation facilitated, however, the development of novel
approaches. Methods based on tensor contraction~\cite{FFC-Compiler}
and symbolic manipulation~\cite{Francis} have been implemented. We
have designed COFFEE to specifically target numerical integration
because it has been demonstrated that it remains the optimal choice
for a wide class of problems~\cite{quadrature1}.

Optimization of local assembly using numerical integration for CPU
platforms has been addressed in FEniCS~\cite{quadrature1}. The
comparison between COFFEE and this work is presented in
Section~\ref{sec:perf-results-fenics}. In~\cite{Markall20101815}, and
more recently in~\cite{petsc-integration-gpu}, the problem has been
studied for GPU architectures. In~\cite{assembly-opencl}, variants of
the standard numerical integration algorithm have been specialized and
evaluated for the PowerXCell processor, but an exhaustive study from
the compiler viewpoint - like ours - is missing, and none of the
optimizations presented in Section~\ref{sec:code-transf} are
mentioned. Among these efforts, to the best of our knowledge, COFFEE
is the first work targeting low-level optimizations through a real
compiler approach.

Our compiler-based optimization approach is made possible by the
top-level DSL, which enables automated code generation. DSLs have been
proven successful in auto-generating optimized code for other domains:
Spiral~\cite{Pueschel:05} for digital signal processing numerical
algorithms, ~\cite{Spampinato:14} for dense linear algebra, or
Pochoir~\cite{pochoir} and SDSL~\cite{stencil-compiler} for image
processing and finite difference stencils. Similarly, PyOP2 is used by
Firedrake to express iteration over unstructured meshes in scientific
codes. COFFEE improves automated code generation in Firedrake.

Many code generators, like those based on the Polyhedral
model~\cite{PLUTO} and those driven by
domain-knowledge~\cite{modeldriven}, make use of cost models. The
alternative of using auto-tuning to select the best implementation for
a given problem on a certain platform has been adopted by
nek5000~\cite{nek5000} for small matrix-matrix multiplies, the ATLAS
library~\cite{ATLAS}, and FFTW~\cite{FFTW} for fast fourier
transforms. In both cases, pruning the implementation space is
fundamental to mitigate complexity and overhead. Likewise, COFFEE uses
a cost model and heuristics (Section~\ref{sec:pyop2-compiler}) to
steer the optimization process.



\section{Conclusions}
\label{sec:conclusions}

In this paper, we have presented the design, optimizations and systematic
performance evaluation of COFFEE, a compiler for finite element local
assembly. In this context, to the best of our knowledge, COFFEE is the
first compiler capable of introducing low-level optimizations to
maximize instruction-level parallelism, register locality and SIMD
vectorization. Assembly kernels have particular characteristics. Their
iteration space is usually very small, with the size depending on
aspects like the degree of accuracy one wants to reach (polynomial
order of the method) and the mesh discretization employed. The data
space, in terms of number of arrays and scalars required to evaluate
the element matrix, grows proportionally with the complexity of the
finite element problem. COFFEE has been developed taking into account
all of these degrees of freedom, based on the idea that reducing the
problem of local assembly optimization to a fixed sequence of
transformations is far too superficial if close-to-peak performance
needs to be reached. The various optimizations overcome limitations of
current vendor and research compilers. The exploitation of domain
knowledge allows some of them to be particularly effective, as
demonstrated by our experiments on two state-of-the-art Intel
platforms. Further work includes a comprehensive study about
feasibility and constraints on transforming kernels into a sequence of
calls to external linear algebra libraries. COFFEE supports all of the
problems expressible in Firedrake, and is already integrated with
this framework.

\bibliographystyle{ACM-Reference-Format-Journals}
\bibliography{biblio}

\medskip

\end{document}